\documentclass{article}
\usepackage{spconf,amsmath,graphicx,hyperref}
\usepackage{booktabs}
\usepackage{cite}
\usepackage{float}
\usepackage{enumitem}
\usepackage{makecell}
\usepackage{xcolor}
\usepackage{pifont}

\newcommand{\cmark}{\ding{51}}  % ✓
\newcommand{\xmark}{\ding{55}}  % ✗

\newcommand{\para}[1]{\vspace{0.5em}\noindent\textbf{#1}}

\title{Evaluating Disentangled Representations for Controllable Music Generation}

\name{Laura Ibáñez-Martínez, Chukwuemeka Nkama, Andrea Poltronieri, Xavier Serra, Martín Rocamora}
\address{Music Technology Group, Universitat Pompeu Fabra, Barcelona, Spain \\
\tt\normalsize\{laura.ibanez, chukwuemeka.nkama, andrea.poltronieri, martin.rocamora\}@upf.edu}

\begin{document}

\maketitle

\begin{abstract}
Recent approaches in music generation rely on disentangled representations,
often labeled as \textit{structure and timbre} or \textit{local and global},
to enable controllable synthesis.
Yet the underlying properties of these embeddings remain underexplored. In this work, we evaluate
such disentangled representations
in a set of
music audio models for controllable generation
using a probing-based framework that goes beyond standard downstream tasks. The selected models reflect
diverse unsupervised disentanglement
strategies, including inductive biases, data augmentations, adversarial objectives, and
staged
training procedures.
We further isolate specific strategies to analyze their effect.
Our analysis spans four key axes: informativeness, equivariance, invariance, and disentanglement, which are assessed across
datasets, tasks, and controlled transformations.
Our findings reveal inconsistencies between
intended and actual semantics of the embeddings, suggesting that current
strategies
fall short of producing truly disentangled representations, and prompting a re-examination of how controllability is approached in music generation.
\end{abstract}


\section{Introduction}
\label{sec:intro}

Over the past few years, text-to-music generation has achieved high-fidelity audio synthesis~\cite{Agostinelli2023, Copet2024, Evans2024}, with research efforts shifting towards improving controllability \cite{Wu2024, Tal2024, Bai2024}.
Although text conditioning provides creative flexibility, it does not enforce temporal
order or
structure~\cite{Agostinelli2023}, and fails to capture lower-level attributes required for precise musical intent~\cite{Wu2024}.
On the other hand, controllability is understood as the ability of a generative model to selectively and predictably manipulate specific attributes of the output (e.g., rhythm or scale)~\cite{Pati2021}.
This directly motivates disentangled representations, commonly described as
those in which each component of learned features refers
to a semantically meaningful concept \cite{Li2018}.

In this context, several methods aim to disentangle music audio into a global embedding,
often
associated with timbre, and a time-varying embedding,
typically linked to structural elements like rhythm or pitch. This separation enables controllable generation and one-shot style transfer~\cite{Cifka2021, Luo2022, Demerle2024}.
Building on earlier work in representation learning that separates pitch and timbre in isolated musical notes~\cite{Luo2019, Luo2020}, recent approaches propose to disentangle these concepts directly from music audio in an unsupervised manner.
This is achieved through architectural constraints~\cite{Cifka2021, Luo2022}, timbre-preserving data augmentations~\cite{Cifka2021, Demerle2024},
adversarial objectives~\cite{Demerle2024}, or two-stage training strategies~\cite{Luo2022, Demerle2024}.
However, the evaluation of these models tends to focus on the generated output rather than on the learned representations themselves, leaving their
suitability for controllable generation
uncertain.
This raises important questions: \textit{What do these representations really capture? Are musical concepts truly disentangled?}

Meanwhile, research on
foundation models
has primarily relied on representation probing as the standard evaluation method for analyzing the information encoded in model embeddings~\cite{Christos2025}.
In the music domain, such frameworks have been applied to evaluate learned representations across several tasks, including key detection, beat tracking, chord recognition, and music tagging~\cite{Won2024, Alonso2025}, as well as to examine how music-theoretic concepts are embedded in the internal representations of generative models~\cite{Wei2024}.
Building on this, Plachouras et al.~\cite{Christos2025} introduced the \texttt{synesis}
framework, originally applied to image and speech models, that assesses representations across multiple axes: \textit{informativeness}, \textit{equivariance}, \textit{invariance}, and \textit{disentanglement}.
Unlike traditional downstream probing, which primarily tests for the presence of task-relevant information in the latent space, this framework examines the structural properties of representations, offering a more holistic
view
of their quality.

In this work, we adapt this framework to the evaluation of disentangled representations in music audio.
Specifically, we: (i) tailor the evaluation protocol to disentangled representations, (ii) apply it to music audio by comparing three unsupervised structure--timbre
disentanglement models, and (iii) investigate the effect of specific disentanglement strategies, such as data augmentations and adversarial objectives, on the learned representations.

Our findings reveal high mutual information between the two embeddings, raising questions about the true semantics of these representations and
their suitability for tasks such as timbre transfer. The results also prompt a broader reflection on the definition of controllability in music generation, suggesting that current unsupervised strategies
fall short of producing representations that
consistently
support it.

\setlength{\tabcolsep}{1.5pt}

\begin{table}[!t]
\resizebox{\columnwidth}{!}{%
\begin{tabular}{l@{\hskip -6pt}ccccccc}
\toprule
\textbf{\textsc{Model}} &
\textbf{\makecell{Timbre \\ Dims}} &
\textbf{\makecell{Struct. \\ Dims}} &
\textbf{\makecell{Temp. \\ Res}} &
\textbf{\makecell{Data \\ Aug.}} &
\textbf{\makecell{Arch. \\ Constr.}} &
\textbf{\makecell{TS \\ Train.}} &
\textbf{\makecell{Adv. \\ Loss}} \\
\midrule
SS-VQ-VAE & 1024 & 1024 & 9 & S+P & \cmark & \xmark & \xmark \\
TS-DSAE   & 16 & 16 & 63 & None & \cmark & \cmark & \xmark \\
AFTER     & 6 & 12 & 86 & S+P+T  & \xmark & \cmark & \cmark \\
\midrule
AFTER-no-aug     & 6 & 12 & 86 & S & \xmark & \cmark & \cmark \\
AFTER-no-adv     & 6 & 12 & 86 & S+P+T & \xmark & \cmark & \xmark  \\
\bottomrule
\end{tabular}
}
\vspace{-6pt}
\caption{
\textbf{Configurations} of evaluated models, including embedding sizes, resolution, and disentanglement strategies.
Data augmentations on the timbre encoder input include segment pairs (S), pitch shifts (P), and time stretches (T).
}
\vspace{-6pt}
\label{tab:disentanglement_methods}
\end{table}

\section{Method}
\label{sec:method}

\subsection{Models and Disentanglement Strategies}
\label{ssec:modelsel}

We assess the following models for
unsupervised disentanglement of music audio into structure and timbre representations:
\begin{itemize}[label={}, leftmargin=*, noitemsep]
    \item \textbf{SS-VQ-VAE} \cite{Cifka2021}, designed for style transfer,
    disentangles \textit{content} and \textit{style} through a discrete codebook for
    content
    and a style encoder regularized with data augmentations.
    \item \textbf{TS-DSAE} \cite{Luo2022} extends the Disentangled Sequential Autoencoder \cite{Li2018} with a two-stage training
    framework
    that
    promotes
    the separation of \textit{local} and \textit{global} factors.
    \item \textbf{AFTER} \cite{Demerle2024}
    aims to disentangle \textit{structure} and \textit{timbre} by combining
    a two-stage training procedure with an adversarial objective
    and timbre-preserving data augmentations.
\end{itemize}

Despite their different terminologies, all models are designed or evaluated under the assumption that global embeddings capture timbre or instrument identity,
while associating the time-varying embedding with structural elements such as pitch,
supporting one-shot timbre transfer.
We therefore adopt a structure-timbre terminology throughout this work.
We
further
use AFTER as a testbed to isolate specific disentanglement strategies, namely data augmentations and adversarial loss,
in order to
study their effect across
evaluation axes.

\subsection{Probing Methods}
\label{ssec:probing}

Probes are classifiers trained to predict a specific property from intermediate representations of a frozen model~\cite{Pilataki2024}.
In recent years, they have become a standard tool for analyzing generative model representations in Music Information Retrieval (MIR)~\cite{Pilataki2024, RodrigoCastellon2021, Wei2024}.
In \texttt{synesis}~\cite{Christos2025}, this corresponds to the axis of \textbf{informativeness}.
Formally, given an embedding $z = f(x)$ and a target label $y$,
probe $g$ is trained to minimize
supervised loss $\mathcal{L}(g(z), y)$, the performance serving as a proxy for the amount of information about $y$ encoded in $z$.

In our setting of disentangled representations, each embedding is probed according to its intended semantics: timbre embeddings on timbre-related tasks such as instrument classification, and structure embeddings on music-theoretic tasks such as multi-pitch estimation and tempo prediction.

\setlength{\tabcolsep}{3pt}

\begin{table}[!t]
\resizebox{\columnwidth}{!}{%
\begin{tabular}{l@{\hskip 0pt}
  c cccc
}
\toprule
&
\multicolumn{1}{c}{\textbf{\textsc{Timbre}}} &
\multicolumn{4}{c}{\textbf{\textsc{Structure}}} \\
\cmidrule(lr){2-2} \cmidrule(lr){3-6}
&
\multicolumn{1}{c}{\textbf{\makecell{S. Instr.}}} &
\multicolumn{1}{c}{\textbf{MPE}} &
\multicolumn{1}{c}{\textbf{\makecell{S. Chords}}} &
\multicolumn{1}{c}{\textbf{\makecell{S. Notes}}} &
\multicolumn{1}{c}{\textbf{\makecell{S. Tempos}}}\\
\cmidrule(lr){2-2} \cmidrule(lr){3-3} \cmidrule(lr){4-4} \cmidrule(lr){5-5}
\cmidrule(lr){6-6}
\textbf{\textsc{Model}} & Acc ↑ & F1 ↑ & Acc ↑ & Acc ↑ & MSE ↓ \\
\midrule
SS-VQ-VAE & \textbf{0.982} & \textbf{0.258} & \textbf{0.462} & \textbf{0.401} & 0.496 \\
TS-DSAE   & 0.286 & 0.133 & 0.243 & 0.354 & \textbf{0.187} \\
AFTER     & 0.284 & 0.162 & 0.263 & 0.311 & 0.745 \\
\midrule
AFTER-no-aug & 0.260 & 0.164* & 0.266* & 0.309 & 0.716* \\
AFTER-no-adv  & 0.266 & 0.168* & 0.251 & 0.280 & 0.794 \\
\bottomrule
\end{tabular}
}
\vspace{-6pt}
\caption{
\textbf{Informativeness} of timbre and structure embeddings
across
datasets and downstream probing tasks.
}
\vspace{-6pt}
\label{tab:informativeness}
\end{table}

\setlength{\tabcolsep}{13pt}

\begin{table*}[ht]
\resizebox{\textwidth}{!}{%
\begin{tabular}{l
  c c c
  c c c
}
\toprule
& \multicolumn{3}{c}{\textbf{\textsc{P-Equivariance}}} &
  \multicolumn{3}{c}{\textbf{\textsc{R-Equivariance}}} \\
\cmidrule(lr){2-4} \cmidrule(lr){5-7}

& \multicolumn{1}{c}{\textbf{\textsc{Timbre}}}
& \multicolumn{2}{c}{\textbf{\textsc{Structure}}}
& \multicolumn{1}{c}{\textbf{\textsc{Timbre}}}
& \multicolumn{2}{c}{\textbf{\textsc{Structure}}} \\
\cmidrule(lr){2-2} \cmidrule(lr){3-4}
\cmidrule(lr){5-5} \cmidrule(lr){6-7}
& \textbf{Instr. Change}
& \textbf{Pitch Shift} & \textbf{Time Stretch}
& \textbf{Instr. Change}
& \textbf{Pitch Shift} & \textbf{Time Stretch} \\
\cmidrule(lr){2-2} \cmidrule(lr){3-3} \cmidrule(lr){4-4}
\cmidrule(lr){5-5} \cmidrule(lr){6-6} \cmidrule(lr){7-7}

\textbf{\textsc{Model}}
& MSE ↓ & MSE ↓ & MSE ↓
& Cos sim ↑ & Cos sim ↑ & Cos sim ↑ \\
\midrule
SS-VQ-VAE &  0.029 & 0.127 & \textbf{0.032} & 0.710 & 0.823 & 0.850 \\
TS-DSAE   & \textbf{0.026} & 0.090 & 0.079 & \textbf{0.838} & \textbf{0.965} & \textbf{0.974} \\
AFTER     & 0.028 & \textbf{0.078} & 0.080 & 0.770 & 0.825 & 0.940 \\
\midrule
AFTER-no-aug & 0.028 & 0.083 & 0.076* & 0.681 & 0.892* & 0.934 \\
AFTER-no-adv  & 0.037 & 0.074* & 0.078* & 0.653 & 0.806 & 0.932 \\
\bottomrule
\end{tabular}
}
\vspace{-6pt}
\caption{
\textbf{P-Equivariance} and \textbf{R-Equivariance} evaluation of timbre and structure embeddings across
defined transformations.
}
\vspace{-6pt}
\label{tab:equivariance}
\end{table*}

\subsection{Beyond Downstream Tasks}

To move beyond downstream prediction, we adopt~\cite{Christos2025}, which assesses representations in terms of their structural response to controlled transformations applied to the input.

\textbf{Equivariance} captures the consistency between transformations in input space and their correspondence in embedding space. Formally, given a transformation $t$ acting on the input $x$, equivariance holds if
$z(t(x)) \approx \rho_{t}(z(x))$,
where $z=f(x)$ is the embedding and $\rho_{t}$ the corresponding action in representation space. We measure this either by predicting transformation parameters (\textit{P-equivariance}) or by predicting the transformed embedding (\textit{R-equivariance}).
\textbf{Invariance}, in contrast, assesses the stability of embeddings under perturbations that should not alter semantic content, measured by the similarity between embeddings of clean and perturbed inputs.

For disentangled representations, transformations are applied selectively: e.g., structure embeddings are evaluated for equivariance to pitch shifts
and invariance to instrument changes, while timbre embeddings are evaluated conversely.

\subsection{Measuring Disentanglement}

\textbf{Disentanglement} is often measured using metrics such as the Mutual Information Gap (MIG)~\cite{Chen2018} or DCI~\cite{Eastwood2018}, which quantify the independence of factors of variation across latent dimensions within a single representation.
Alternatively, Plachouras et al.~\cite{Christos2025} proposed to assess disentanglement
by measuring probe performance drops when predicting one factor under controlled transformations of another.

In our setting, where timbre and structure are encoded separately, we extend this principle by comparing
probe performance when trained on a single embedding versus on the concatenation of both.
Formally, for a task associated with embedding $z_t$, we compute:
\[
\Delta = \big| \text{Acc}(g'(z_t \oplus z_s)) - \text{Acc}(g(z_t)) \big|,
\]
where $\text{Acc}$ denotes probe accuracy, $g$ is the corresponding informativeness probe
trained on $z_t$, and $g'$ is a new probe trained on the concatenation ($\oplus$) of $z_t$ and $z_s$.
A large $\Delta$ indicates that the complementary embedding carries residual information about the task, suggesting entanglement, while values closer to zero imply better
embedding separation.

\setlength{\tabcolsep}{6pt}

\begin{table}[!t]
\resizebox{\columnwidth}{!}{%
\begin{tabular}{l c c c}
\toprule
&
\multicolumn{2}{c}{\textbf{\textsc{Timbre}}} &
\multicolumn{1}{c}{\textbf{\textsc{Structure}}} \\
\cmidrule(lr){2-3} \cmidrule(lr){4-4}
& \textbf{Pitch Shift} & \textbf{Time Stretch}
& \textbf{Instr. Change} \\
\cmidrule(lr){2-2} \cmidrule(lr){3-3} \cmidrule(lr){4-4}

\textbf{\textsc{Model}}
& Cos sim ↑ & Cos sim ↑
& Cos sim ↑ \\
\midrule
SS-VQ-VAE & \textbf{0.667} & 0.963 & 0.919 \\
TS-DSAE   & 0.491 & 0.993 & \textbf{0.960} \\
AFTER     & 0.546 & \textbf{0.996} & \textbf{0.960} \\
\midrule
AFTER-no-aug     & 0.576* & 0.997* & 0.955 \\
AFTER-no-adv     & 0.366 & 0.984 & 0.925 \\
\bottomrule
\end{tabular}
}
\vspace{-6pt}
\caption{
\textbf{Invariance} of timbre and structure embeddings
under
different perturbations, reported as cosine similarity.
}
\vspace{-6pt}
\label{tab:invariance}
\end{table}

\section{Experimental Setup}
\label{sec:expsetup}

\subsection{Model Training and Configurations}
\label{ssec:modeltraining}

For a fair comparison between the different models and disentanglement strategies, we retrain all variants using the Slakh2100 \cite{Manilow2019} dataset, a 145-hour collection of
synthetic mixtures with
corresponding individual stems.
As in \cite{Demerle2024}, mixes and drums are excluded, while
the
remaining
stems are divided between
90\% for training and 10\% for validation.

The three models SS-VQ-VAE, TS-DSAE, and AFTER
are trained using their official repositories\footnote{For AFTER, we rely on the real-time model implementation.}
with default configurations,
assumed to be optimal for each disentanglement strategy despite differing embedding sizes.
In addition, we train two variants of AFTER to isolate specific disentanglement strategies: one
without pitch and tempo augmentations
applied to the timbre encoder input
(\textit{AFTER-no-aug}), and one without the adversarial loss (\textit{AFTER-no-adv}).
Table~\ref{tab:disentanglement_methods} summarizes the different
model configurations
under study.

\subsection{Evaluation Framework}
\label{ssec:evalframework}

We build on the \texttt{synesis} evaluation framework~\cite{Christos2025}, adapting it\footnote{
Additional implementation details and an overview figure
are available at: \url{https://github.com/lauraibnz/synesis}} to our setting
through
the following components.

\para{Probing datasets.}
We primarily rely on the \textit{SynTheory (S.)} dataset~\cite{Pilataki2024},
which provides controlled synthesized variations
defined over
Western music-theoretic
concepts,
expected to
be present
in Slakh2100 in a more complex and naturally entangled form.
For timbre,
we
use
\textit{SynTheory Instruments},
derived from
the
\textit{Chords} subset
by performing instrument classification
(92
MIDI program numbers).
For
structure, we use \textit{SynTheory Notes}
for
root-note classification (12 pitch classes), \textit{SynTheory Tempos} for tempo regression (161 bpm values), and \textit{SynTheory Chords} for chord
classification
(4 chord types).
We additionally evaluate the structure embedding on multi-pitch estimation (MPE)
to measure its performance on note-level tasks.
For this, we
train on
4-second excerpts from 36 hours of MAESTRO data~\cite{Hawthorne2019},
and evaluate on the MusicNet test set with corrected alignments~\cite{Maman2022}.

\para{Factors of variation.}
We consider pitch, tempo, and instrument
timbre
as factors of variation, and probe equivariance and invariance
using
three corresponding transformations.
\textit{Instrument Change},
derived from
SynTheory,
uses identical structures rendered
with different instruments and is quantified using the timbre dissimilarity metric of~\cite{Cifka2021}, which measures perceptual distance between instrument timbres. This tests timbre equivariance and structure invariance.
\textit{Pitch Shift} and \textit{Time Stretch}, implemented in \texttt{synesis}, parameterize pitch and tempo variations and are used to evaluate structure equivariance and timbre invariance.

\para{Training details.}
For standard classification and regression tasks, we follow
\texttt{synesis}
and train shallow linear probes (SLPs)
with early stopping on the frozen embeddings
across all evaluation axes.
For global tasks on the structure embedding, we apply
average
pooling over the time dimension. For multi-pitch estimation, we use a two-layer MLP with 512 hidden units and apply a sigmoid activation at the output layer.

\para{Evaluation metrics.}
Across all evaluation axes, we adopt the default metrics from \texttt{synesis}: accuracy for classification, mean squared error (MSE) for regression, and cosine similarity for embedding similarity. For multi-pitch estimation, we use the
\texttt{mir\_eval} library~\cite{Raffel2014} to compute F1 scores. The decision threshold is tuned on the validation set and then applied to the test set, with final results averaged at the track level.

\setlength{\tabcolsep}{2pt}

\begin{table}[!t]
\resizebox{\columnwidth}{!}{%
\begin{tabular}{l@{\hskip 0pt}
  c cccc
}
\toprule
&
\multicolumn{1}{c}{\textbf{\textsc{Timbre}}} &
\multicolumn{4}{c}{\textbf{\textsc{Structure}}} \\
\cmidrule(lr){2-2} \cmidrule(lr){3-6}

& \multicolumn{1}{c}{\textbf{\makecell{S. Instr.}}} &
\multicolumn{1}{c}{\textbf{MPE}} &
\multicolumn{1}{c}{\textbf{\makecell{S. Chords}}} &
\multicolumn{1}{c}{\textbf{\makecell{S. Notes}}} &
\multicolumn{1}{c}{\textbf{\makecell{S. Tempos}}}\\
\cmidrule(lr){2-2} \cmidrule(lr){3-3} \cmidrule(lr){4-4}
\cmidrule(lr){5-5} \cmidrule(lr){6-6}

\textbf{\textsc{Model}} & $\Delta$Acc ↓
& $\Delta$F1 ↓
& $\Delta$Acc ↓
& $\Delta$Acc ↓
& $\Delta$MSE ↓ \\
\midrule
SS-VQ-VAE & \textbf{0.002} & 0.031  & 0.311 & 0.318 & 0.478  \\
TS-DSAE   & 0.015 & 0.016 & 0.066 & 0.034 & \textbf{0.174} \\
AFTER     & 0.068 & \textbf{0.005} & \textbf{0.001} & \textbf{0.009} & 0.382 \\
\midrule
AFTER-no-aug     & 0.097 & 0.003*  & 0.048 & 0.004* & 0.458 \\
AFTER-no-adv     & 0.151 & 0.056 & 0.067 & 0.015 & 0.298*  \\
\bottomrule
\end{tabular}
}
\vspace{-6pt}
\caption{\textbf{Disentanglement} evaluation across downstream tasks. $\Delta$ measures
difference
from concatenated embeddings.}
\vspace{-6pt}
\label{tab:disentanglement}
\end{table}

\section{Results}
\label{sec:results}

In all tables, best results are in \textbf{bold}, and asterisk (*)  indicates an AFTER variant outperforming the
default configuration.

\para{Informativeness.}
As shown in Table~\ref{tab:informativeness}, \textit{SS-VQ-VAE} performs best on instrument classification for timbre, likely due to its larger embedding size. This aligns with~\cite{Christos2025}, which links informativeness to embedding capacity, and the same trend extends to most structure tasks.
Remarkably, it also achieves the best multi-pitch estimation performance despite its lower temporal resolution; however,
overall
performance on this note-level task remains low. In contrast, \textit{TS-DSAE}
is most effective for tempo regression,
suggesting that its architecture and training strategy compensate for its smaller embedding size.
\textit{AFTER} remains
less consistent,
with notably weaker tempo prediction and only minor variant-specific effects.

\para{Equivariance.}
\label{ssec:equivarianceeval}
Table~\ref{tab:equivariance} summarizes the equivariance results,
revealing
a negative correlation with informativeness and embedding size:
\textit{SS-VQ-VAE},
while highly informative, is less equivariant.
This trade-off is consistent with
observations in the original framework \cite{Christos2025}. In contrast, \textit{TS-DSAE}, despite its compact structure embedding, achieves the strongest R-equivariance across all transformations.
\textit{AFTER} attains intermediate performance, with variants showing reduced
timbre equivariance
and mixed outcomes for structure.

\para{Invariance.}
\label{ssec:invarianceeval}
Table \ref{tab:invariance} presents the invariance results. Timbre embeddings are robust to tempo perturbations but sensitive to pitch shifts. \textit{SS-VQ-VAE} shows the strongest pitch invariance, followed by \textit{AFTER}. Its variant without adversarial loss is most vulnerable, while removing pitch and tempo augmentations
surprisingly results in higher invariance
to both factors.
Structure embeddings are broadly invariant to instrument changes, though \textit{AFTER} again weakens when adversarial loss is removed. Consistent with the original framework \cite{Christos2025}, invariance appears driven less by
capacity
and more by training objectives and disentanglement strategies.

\para{Disentanglement.}
\label{ssec:disentanglementeval}
Table~\ref{tab:disentanglement} summarizes the disentanglement results. \textit{SS-VQ-VAE} shows little timbre leakage into the structure embedding, but its timbre embedding carries substantial structural information, likely due to its
higher capacity
and despite data augmentations. \textit{AFTER}, in contrast, reveals the opposite trend: its structure embedding contains notable timbre information, while the timbre embedding generally
carries
little pitch information.
Additional
structural leakage is
observed in variants without disentanglement strategies. Across models, timbre embeddings systematically encode tempo, as concatenation consistently improves tempo regression performance. Overall, \textit{TS-DSAE}
appears
the most balanced, with concatenation
having
only marginal effects.

\section{Conclusion}
\label{sec:conclusion}
We presented a systematic evaluation of disentangled music representations, adapting a recent framework~\cite{Christos2025} that
goes
beyond standard downstream probing. Our analysis assessed informativeness, equivariance, invariance, and disentanglement across modern unsupervised architectures for structure--timbre disentanglement in music audio, while also
investigating the role of specific disentanglement strategies.

Results show systematic inconsistencies between the intended and actual semantics of the embeddings. We observe asymmetric leakage between timbre and structure,
as well as
persistent tempo
information
encoded
in
timbre embeddings.
This limits controllability and reduces
suitability for timbre transfer tasks.
Moreover, while these models achieve high reconstruction quality,
note-level tasks with
simple
probes fail to extract
sufficient
information from the embeddings, suggesting
such
information is
encoded in a highly complex
way.

More broadly, our evaluation confirms trends observed in the original framework: a trade-off between informativeness and equivariance, with larger embeddings encoding more information but being less transformation-consistent.
At the same time, disentanglement strategies such as adversarial losses or data augmentations have the strongest impact on invariance and disentanglement, influencing how well factors are separated rather than how much information is captured.

Finally, while our focus has been on representation-level analysis, controllability
also depends on decoding strategies.
As noted in~\cite{Pati2021}, disentangled representations are necessary but not sufficient:
predictable control
requires
decoders to
preserve factor separation at the signal level.
Extending representation-level diagnostics with output-level evaluations, including controlled generation experiments and perceptual listening-based assessments of controllability,
remains an important direction for future work.

\section{Acknowledgments}

This work has been supported by
\textit{IA y Música: Cátedra en Inteligencia Artificial y Música} (TSI-100929-2023-1), funded by the Secretaría de Estado de Digitalización e Inteligencia Artificial and the European Union-Next Generation EU, and 
\textit{IMPA: Multimodal AI for Audio Processing} (PID2023-152250OB-I00), funded by the Ministry of Science, Innovation and Universities of the Spanish Government, the Agencia Estatal de Investigación (AEI) and co-financed by the European Union.

\bibliographystyle{IEEEbib}
\bibliography{ICASSP_disentanglement}

@Misc{Agostinelli2023,
  author = {Agostinelli, Andrea and Denk, Timo I. and Borsos, Zalán and Engel, Jesse and Verzetti, Mauro and Caillon, Antoine and Huang, Qingqing and Jansen, Aren and Roberts, Adam and Tagliasacchi, Marco and others},
  title  = {{MusicLM}: {Generating Music From Text}},
  year   = {2023},
}

@InProceedings{Cifka2021,
  author     = {Cífka, Ondřej and Ozerov, Alexey and Şimşekli, Umut and Richard, Gaël},
  booktitle  = {Proc. ICASSP},
  title      = {{Self-Supervised {VQ}-{VAE} for One-Shot Music Style Transfer}},
  year       = {2021},
  pages      = {96--100},
  doi        = {10.1109/ICASSP39728.2021.9414235},
  eventtitle = {{ICASSP} 2021 - 2021 {IEEE} International Conference on Acoustics, Speech and Signal Processing ({ICASSP})},
  url        = {https://ieeexplore.ieee.org/document/9414235},
  urldate    = {2025-03-26},
}

@Article{Copet2024,
  author  = {Copet, Jade and Kreuk, Felix and Gat, Itai and Remez, Tal and Kant, David and Synnaeve, Gabriel and Adi, Yossi and D{\'e}fossez, Alexandre},
  journal = {Proc. NeurIPS},
  title   = {{Simple and Controllable Music Generation}},
  year    = {2024},
  volume  = {36},
}

@InProceedings{Demerle2024,
  author    = {Demerlé, Nils and Esling, Philippe and Doras, Guillaume and Genova, David},
  booktitle = {Proc. ISMIR},
  title     = {Combining audio control and style transfer using latent diffusion},
  year      = {2024},
  doi       = {10.48550/arXiv.2408.00196},
}

@InProceedings{Evans2024,
  author    = {Evans, Zach and Parker, Julian D. and Carr, CJ and Zukowski, Zack and Taylor, Josiah and Pons, Jordi},
  booktitle = {Proc. ICASSP},
  title     = {{Stable Audio Open}},
  year      = {2025},
  pages     = {1-5},
  doi       = {10.1109/ICASSP49660.2025.10888461},
}

@InProceedings{Hawthorne2019,
  author    = {Curtis Hawthorne and Andriy Stasyuk and Adam Roberts and Ian Simon and Cheng-Zhi Anna Huang and Sander Dieleman and Erich Elsen and Jesse Engel and Douglas Eck},
  booktitle = {Proc. ICLR},
  title     = {{Enabling Factorized Piano Music Modeling and Generation with the {MAESTRO} Dataset}},
  year      = {2019},
  url       = {https://openreview.net/forum?id=r1lYRjC9F7},
}

@InProceedings{Luo2022,
  author    = {Luo, Yin-Jyun and Ewert, Sebastian and Dixon, Simon},
  booktitle = {Proc. IJCAI},
  title     = {{Towards Robust Unsupervised Disentanglement of Sequential Data — A Case Study Using Music Audio}},
  year      = {2022},
  pages     = {3299--3305},
  doi       = {10.24963/ijcai.2022/458},
  url       = {https://doi.org/10.24963/ijcai.2022/458},
}

@InProceedings{Maman2022,
  author    = {Maman, Ben and Bermano, Amit H},
  booktitle = {Proc. ICML},
  title     = {{Unaligned Supervision for Automatic Music Transcription in The Wild}},
  year      = {2022},
  pages     = {14918--14934},
  url       = {https://proceedings.mlr.press/v162/maman22a.html},
}

@InProceedings{Manilow2019,
  author    = {Manilow, Ethan and Wichern, Gordon and Seetharaman, Prem and Le Roux, Jonathan},
  booktitle = {Proc. IEEE WASPAA},
  title     = {{Cutting Music Source Separation Some Slakh: A Dataset to Study the Impact of Training Data Quality and Quantity}},
  year      = {2019},
  pages     = {45--49},
}

@InProceedings{Pilataki2024,
  author    = {Pilataki, Mary and Mauch, Matthias and Dixon, Simon},
  booktitle = {Proc. ACM MM Asia},
  title     = {{Pitch-aware Generative Pretraining Improves Multi-Pitch Estimation with Scarce Data}},
  year      = {2024},
  articleno = {41},
  doi       = {10.1145/3696409.3700202},
  isbn      = {9798400712739},
  numpages  = {8},
  url       = {https://doi.org/10.1145/3696409.3700202},
}

@InProceedings{Raffel2014,
  author    = {Colin Raffel and Brian McFee and Eric J. Humphrey and Justin Salamon and Oriol Nieto and Dawen Liang and Daniel P. W. Ellis},
  booktitle = {Proc. ISMIR},
  title     = {mir\_eval: A {Transparent} {Implementation} {of} {Common} {MIR} {Metrics}},
  year      = {2014},
  pages     = {367--372},
  url       = {https://colinraffel.com/publications/ismir2014mir_eval.pdf},
  urldate   = {2025-03-25},
}

@InProceedings{RodrigoCastellon2021,
  author     = {Rodrigo Castellon and Chris Donahue and Percy Liang},
  booktitle  = {Proc. ISMIR},
  title      = {Codified audio language modeling learns useful representations for music information retrieval},
  year       = {2021},
  date       = {2021},
  eventtitle = {International Society for Music Information Retrieval},
  url        = {https://archives.ismir.net/ismir2021/paper/000010.pdf},
  urldate    = {2025-03-27},
}

@InProceedings{Wei2024,
  author    = {Megan Wei and Michael Freeman and Chris Donahue and Chen Sun},
  booktitle = {Proc. ISMIR},
  title     = {{Do Music Generation Models Encode Music Theory?}},
  year      = {2024},
  pages     = {680-687},
}

@InProceedings{Christos2025,
  author    = {Christos Plachouras and Julien Guinot and George Fazekas and Elio Quinton and Emmanouil Benetos and Johan Pauwels},
  booktitle = {Proc. IJCNN},
  title     = {{Towards a Unified Representation Evaluation Framework Beyond Downstream Tasks}},
  year      = {2025},
}

@Article{Wu2024,
  author     = {Wu, Shih-Lun and Donahue, Chris and Watanabe, Shinji and Bryan, Nicholas J.},
  journal    = {IEEE/ACM Trans. Audio, Speech, Lang. Process.},
  title      = {{Music ControlNet}: {Multiple Time-Varying Controls for Music Generation}},
  year       = {2024},
  issn       = {2329-9290},
  month      = may,
  pages      = {2692–2703},
  volume     = {32},
  doi        = {10.1109/TASLP.2024.3399026},
  issue_date = {2024},
  numpages   = {12},
  publisher  = {IEEE Press},
  url        = {https://doi.org/10.1109/TASLP.2024.3399026},
}

@InProceedings{Tal2024,
  author    = {Or Tal and Alon Ziv and Itai Gat and Felix Kreuk and Yossi Adi},
  booktitle = {Proc. ISMIR},
  title     = {{Joint Audio and Symbolic Conditioning for Temporally Controlled Text-to-Music Generation}},
  year      = {2024},
  pages     = {264-271},
  doi       = {10.5281/zenodo.14877325},
  url       = {https://doi.org/10.5281/zenodo.14877325},
  venue     = {San Francisco, California, USA and Online},
}

@Misc{Bai2024,
  author        = {Ye Bai and Haonan Chen and Jitong Chen and Zhuo Chen and Yi Deng and Xiaohong Dong and Lamtharn Hantrakul and Weituo Hao and Qingqing Huang and Zhongyi Huang and others},
  title         = {{Seed-Music}: {A Unified Framework for High Quality and Controlled Music Generation}},
  year          = {2024},
  archiveprefix = {arXiv},
  eprint        = {2409.09214},
  primaryclass  = {cs.SD},
  url           = {https://arxiv.org/abs/2409.09214},
}

@Article{Luo2020,
  author  = {Luo, Yin-Jyun and Cheuk, Kin Wai and Nakano, Tomoyasu and Goto, Masataka and Herremans, Dorien},
  journal = {Proc. ISMIR},
  title   = {{Unsupervised Disentanglement of Pitch and Timbre for Isolated Musical Instrument Sounds}},
  year    = {2020},
}

@InProceedings{Luo2019,
  author    = {Yin-Jyun Luo and Kat Agres and Dorien Herremans},
  booktitle = {Proc. ISMIR},
  title     = {{Learning Disentangled Representations of Timbre and Pitch for Musical Instrument Sounds Using Gaussian Mixture Variational Autoencoders}},
  year      = {2019},
  pages     = {746-753},
}

@InProceedings{Won2024,
  author    = {Won, Minz and Hung, Yun-Ning and Le, Duc},
  booktitle = {Proc. ICASSP},
  title     = {{A Foundation Model for Music Informatics}},
  year      = {2024},
  pages     = {1226-1230},
  doi       = {10.1109/ICASSP48485.2024.10448314},
}

@InProceedings{Pati2021,
  author    = {Ashis Pati and Alexander Lerch},
  booktitle = {Proc. ISMIR},
  title     = {{Is Disentanglement enough? On Latent Representations for Controllable Music Generation}},
  year      = {2021},
  pages     = {517-524},
}

@InProceedings{Li2018,
  author    = {Yingzhen Li and Stephan Mandt},
  booktitle = {Proc. ICML},
  title     = {{Disentangled Sequential Autoencoder}},
  year      = {2018},
  url       = {https://api.semanticscholar.org/CorpusID:48353305},
}

@InProceedings{Alonso2025,
  author    = {Alonso-Jim{\'e}nez, Pablo and Ramoneda, Pedro and Araz, R. Oguz and Poltronieri, Andrea and Bogdanov, Dmitry},
  booktitle = {Proc. ACM MM},
  title     = {{OMAR-RQ}: {Open Music Audio Representation Model Trained with Multi-Feature Masked Token Prediction}},
  year      = {2025},
  doi       = {10.1145/3746027.3756871},
}

@InProceedings{Chen2018,
  author    = {Chen, Ricky T. Q. and Li, Xuechen and Grosse, Roger and Duvenaud, David},
  booktitle = {Proc. NIPS},
  title     = {{Isolating Sources of Disentanglement in VAEs}},
  year      = {2018},
  pages     = {2615–2625},
  location  = {Montr\'{e}al, Canada},
  numpages  = {11},
}

@InProceedings{Eastwood2018,
  author    = {Cian Eastwood and Christopher K. I. Williams},
  booktitle = {Proc. ICLR},
  title     = {{A Framework for the Quantitative Evaluation of Disentangled Representations}},
  year      = {2018},
  url       = {https://openreview.net/forum?id=By-7dz-AZ},
}

\end{document}